\documentclass[12pt]{iopart}
\usepackage{graphicx}
\newcommand{\Msun}{$\mathrm{M}_{\odot}$}
\begin{document}

\title{Constraining properties of neutron stars with heavy-ion reactions in terrestrial laboratories}
\author{Bao-An Li$^1$, Lie-Wen Chen$^2$, Che Ming Ko$^3$, Plamen~G.~Krastev$^1$, Andrew W. Steiner$^4$, Gao-Chan Yong$^1$}
\address{$^1$Department of Physics, Texas A\&M University-Commerce, Commerce, TX 75429, U.S.A}
\address{$^2$Institute of Theoretical Physics, Shanghai Jiao Tong University, Shanghai 200240}
\address{$^3$Cyclotron Institute and Physics Department, Texas A\&M University,
College Station, Texas 77843, USA}
\address{$^4$Joint Institute for Nuclear Astrophysics and National Superconducting Cyclotron Laboratory, Michigan State University, East Lansing, MI 48824, USA}

\begin{abstract}
Heavy-ion reactions provide a unique means to investigate the
equation of state (EOS) of neutron-rich nuclear matter, especially
the density dependence of the nuclear symmetry energy
$E_{sym}(\rho)$. The latter plays an important role in understanding
many key issues in both nuclear physics and astrophysics. Recent
analyses of heavy-ion reactions have already put a stringent constraint
on the $E_{sym}(\rho)$ around the saturation density. This subsequently
allowed us to constrain significantly the radii and cooling mechanisms of neutron stars as
well as the possible changing rate of the gravitational constant G.
\end{abstract}

\section{Introduction}
Within the parabolic approximation which has been verified by all
many-body theories to date, the EOS of isospin asymmetric nuclear
matter can be written as $E(\rho ,\delta )=E(\rho ,\delta =0)+E_{\rm
sym}(\rho )\delta ^{2} +\mathcal{O}(\delta^4)$, where
$\delta\equiv(\rho_{n}-\rho _{p})/(\rho _{p}+\rho _{n})$ is the
isospin asymmetry and $E_{\rm sym}(\rho)$ is the density-dependent
nuclear symmetry energy. The latter is very important for
understanding many interesting astrophysical
problems~\cite{lat01,steiner05}, the structure of radioactive
nuclei~\cite{brown,stone}, and the dynamics of heavy-ion
reactions~\cite{ireview98,ibook01,dan02,ditoro}.
\section{Constraining the density dependence of the nuclear symmetry energy
in terrestrial nuclear laboratories}
Considerable progress has been
made recently in determining the density dependence of the nuclear
symmetry energy at sub-normal densities in terrestrial nuclear
laboratories.
\begin{figure}[htb]
\begin{center}
\includegraphics[width=6.cm,height=5.cm]{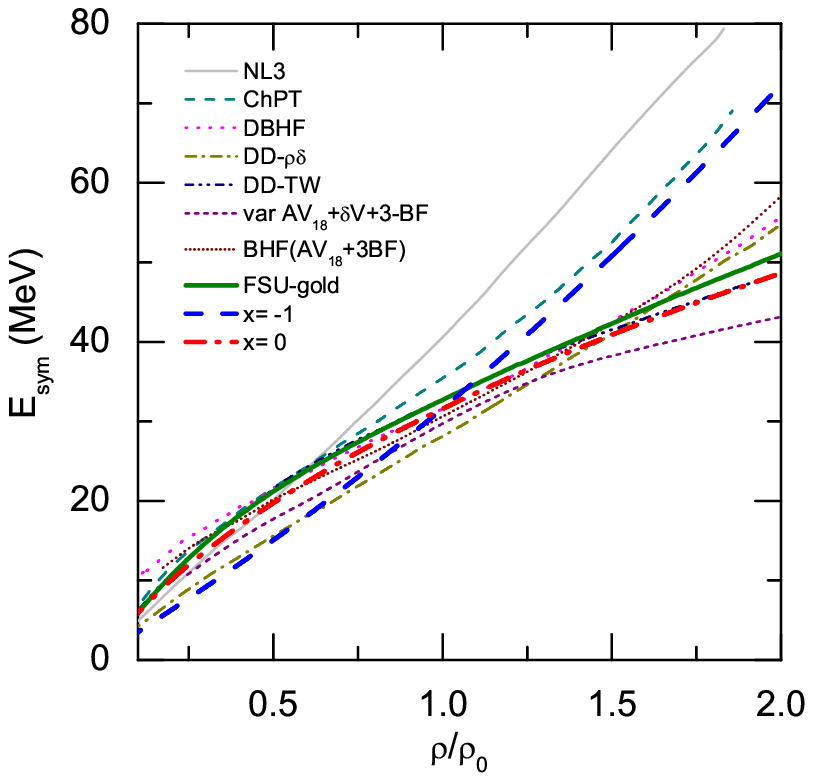}
\includegraphics[width=6cm,height=5.cm]{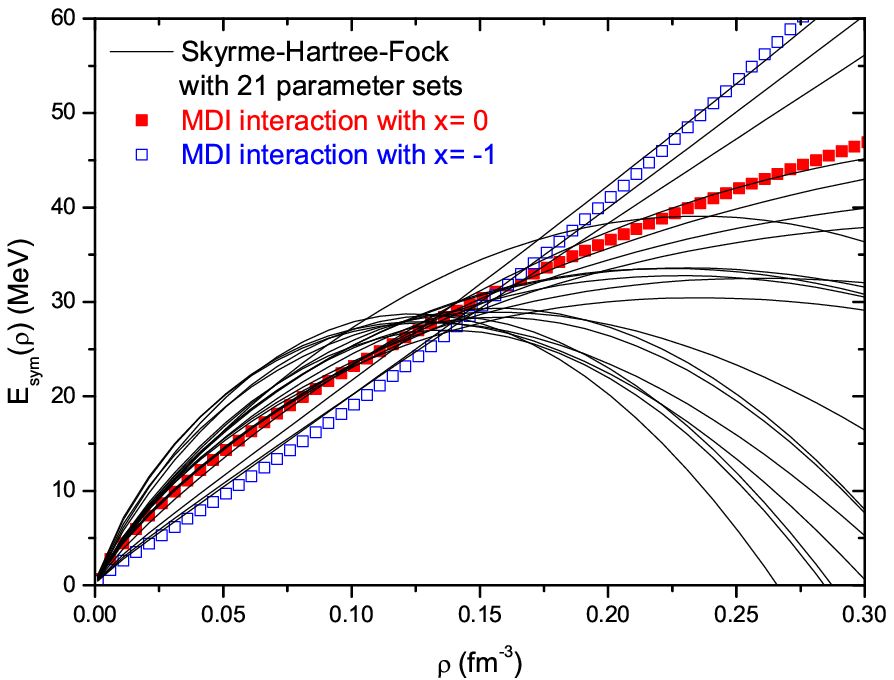}
\caption{{\protect\small Left panel:Density dependence of the
nuclear symmetry energy using the MDI interaction with $x=0$ and
$x=-1$ and other many-body theories predictions (taken from
\protect\cite{gao07}). Right panel: symmetry energies obtained from
21 sets of Skyrme interactions and the MDI interaction with $x=-1$
and $x=0$~\cite{chenkl}.}} \label{esym}
\end{center}
\end{figure}
Shown in Fig.~\ref{esym} are several typical theoretical model
predictions \cite{diep,fuchs06,zuo02} compared with the constraints
obtained from model analyses of experimental data (those labeled
$x=0$, $x=-1$ and FSU-Gold). The constraints labeled $x=0$ and
$x=-1$ were extracted recently from studying isospin diffusion in
the reaction of $^{124}$Sn +$^{112}$Sn at $E_{beam}/A=50$ MeV within
a transport model \cite{ls03,mbt,chen04,li05} using the MDI
interaction \cite{das03}. For this particular reaction the maximum
density reached is about $1.2\rho_0$. The constraints are thus only
valid below this density. Moreover, it was shown that the
neutron-skin thickness in $^{208}$Pb calculated within the
Hartree-Fock approach using the same underlying Skyrme interactions
as the ones labeled $x=0$ and $x=-1$ is consistent with the
available experimental data \cite{steiner05b,ba0511,chenkl}. The
symmetry energy labeled as FSU-Gold was calculated within a
Relativistic Mean Field Model (RMF) using a parameter set such that
it reproduces both the giant monopole resonance in $^{90}$Zr and
$^{208}$Pb, and the isovector giant dipole resonance of $^{208}$Pb
\cite{piek05}. These results all together represent the best
phenomenological constraints available on the symmetry energy at
sub-normal densities. The various predictions at supra-normal
densities still diverge widely. Hopefully, nuclear reactions with
high energy radioactive beams will allow us to pin down the high
density behavior soon using several probes predicted
recently \cite{chen07}.

The available constraints on the symmetry energy limit the nuclear
effective interactions in nuclear matter. This can be seen by
comparing them with the symmetry energies obtained from Skyrme
effective interactions~\cite{chenkl}.  The right panel of Fig.
\ref{esym} displays the density dependence of $E_{\rm sym}(\rho )$
for $21$ sets of Skyrme interaction parameters that are currently
used widely in nuclear structure studies. Surprisingly, most of
these effective interactions lead to symmetry energies rather
inconsistent with the constraints discussed above.

\section{Constraining the radii and cooling mechanisms of neutron stars}
While the maximum mass of neutron stars is mainly determined by the
incompressibility of symmetric nuclear matter, their radii are
primarily determined by the isospin asymmetric pressure that is
proportional to the slope of the symmetry energy $E_{\rm
sym}^{\prime}(\rho)$. For the simplest case of a
neutron-proton-electron ($npe$) matter in neutron stars at $\beta$
equilibrium, the pressure is given by $
P(\rho,\delta)=P_0(\rho)+P_{\rm asy}(\rho,\delta)=
\rho^2\left(\frac{\partial E}{\partial \rho}
\right)_{\delta}+\frac{1}{4}\rho_e\mu_e\nonumber\\
=\rho^2\left[E'(\rho,\delta=0)+E'_{\rm sym}(\rho)\delta^2\right]
+\frac{1}{2}\delta(1-\delta)\rho E_{\rm sym}(\rho), $ where the
electron density is $\rho_e=\frac{1}{2}(1-\delta)\rho$ and the
chemical potential is $\mu_e=\mu_n-\mu_p=4\delta E_{\rm sym}(\rho)$.
The equilibrium value of $\delta$ is determined by the chemical
equilibrium and charge neutrality conditions, i.e., $\delta=1-2x_p$
with $ x_p\approx 0.048 \left[E_{\rm sym}(\rho)/E_{\rm
sym}(\rho_0)\right]^3 (\rho/\rho_0)(1-2x_p)^3.$ Because of the large
$\delta$ value in neutron stars, the electron degenerate pressure is
small. Moreover, the isospin symmetric contribution to the pressure
is also very small around normal nuclear matter density as
$E'(\rho_0,\delta=0)=0$. The pressure is thus dominated by the term
proportional to the slope of the symmetry energy. Since neutron star
radii are determined by the pressure at moderate densities where the
proton content of matter is small, they are very sensitive to the
slope of the symmetry energy near and just above $\rho_0$. In
particular, a stiffer symmetry energy is expected to lead to a
larger neutron star radius.
\begin{figure}[th]
\begin{center}
\includegraphics[width=5.cm,height=5.cm]{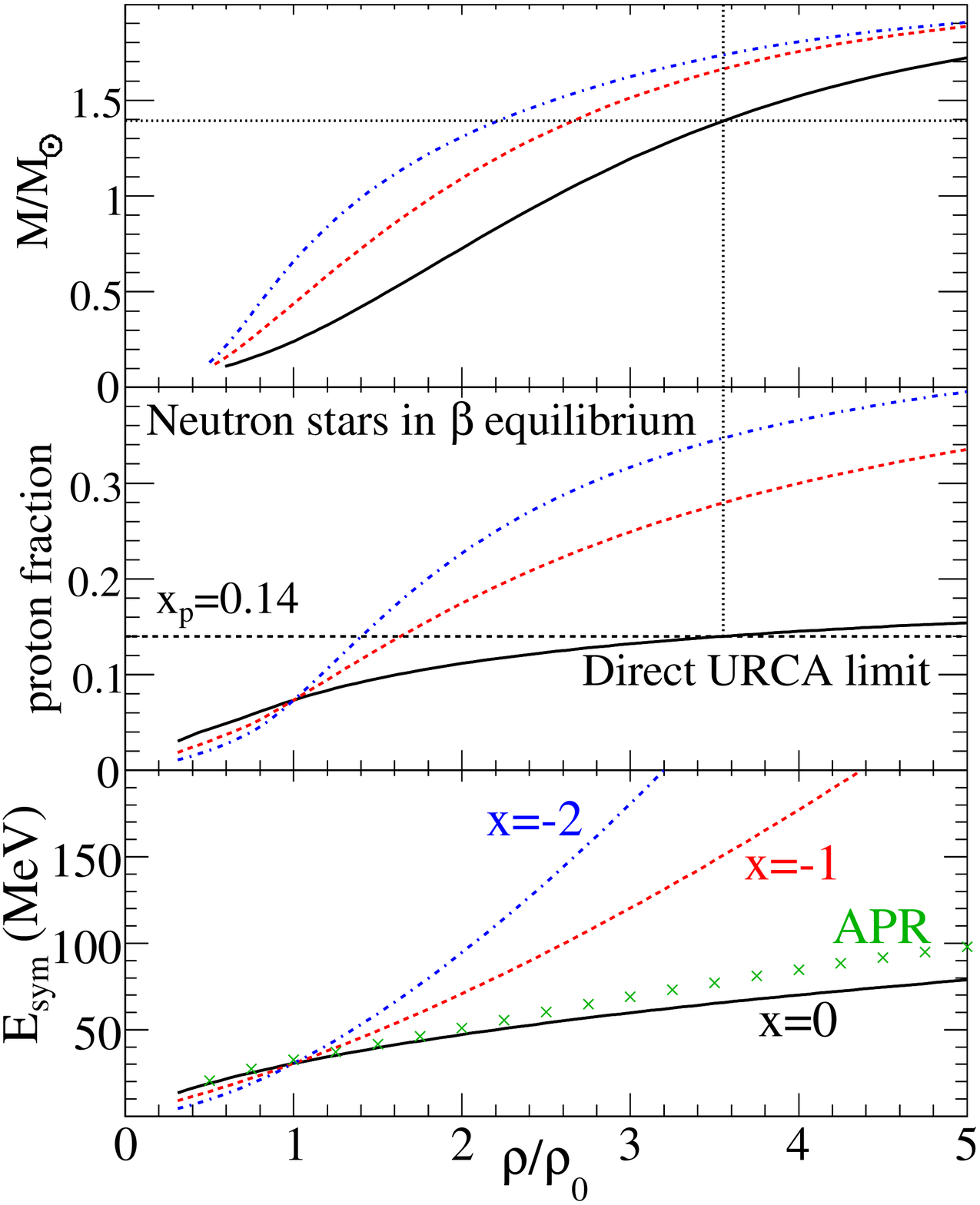}
\includegraphics[width=5.cm,height=5.cm]{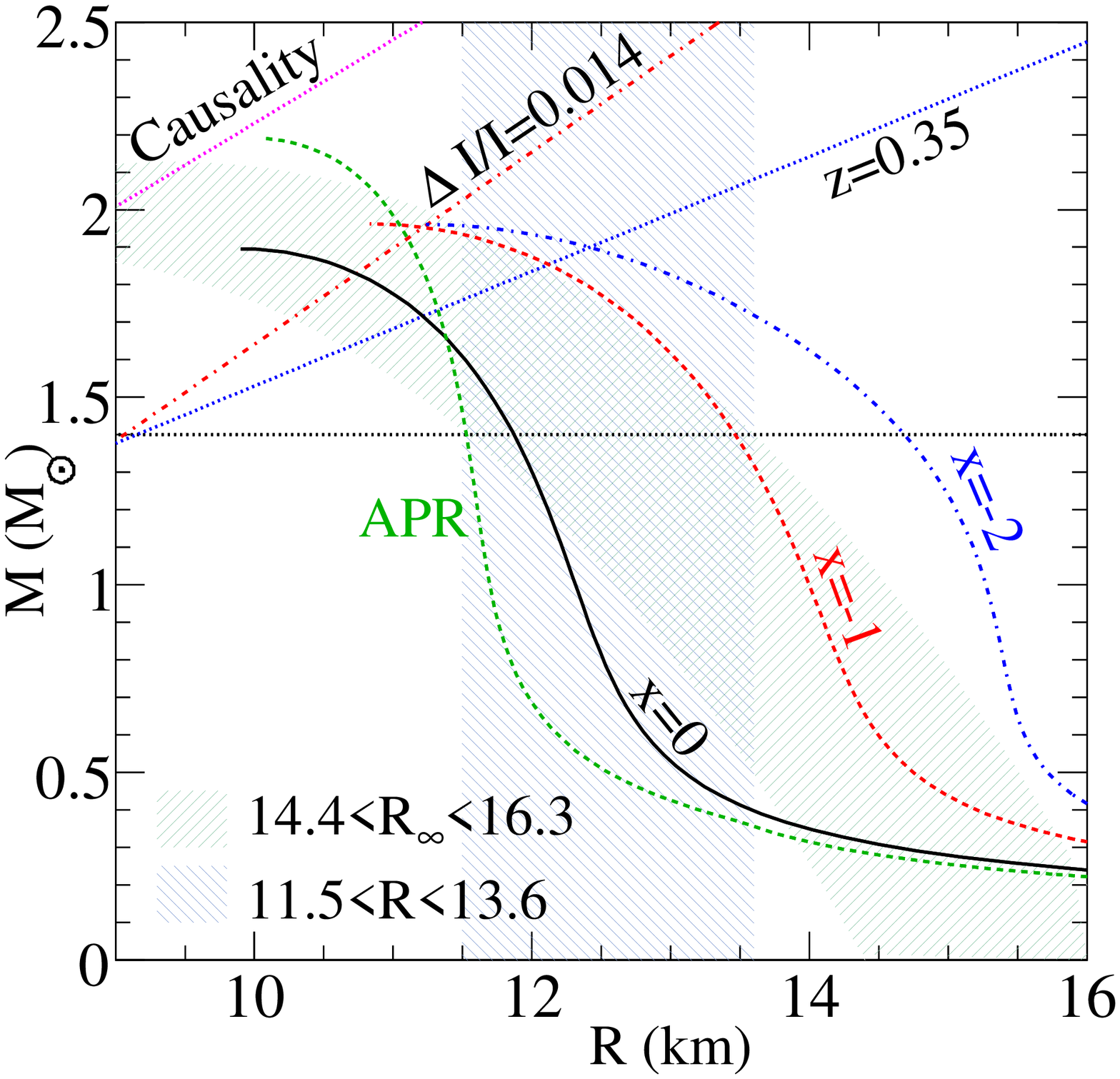}
\caption{{\protect\small Left panel: mass and proton fraction of
neutron star and the symmetry energy as functions of density. Right
panel: correlations in neutron star masses and radii~\cite{ba0511}.
All results are for spherically-symmetric, non-rotating,
non-magnetized neutron stars consisting of $npe\mu$ matter at zero
temperature.}}
\end{center}
\label{mr}
\end{figure}

The dependence of some basic properties of neutron stars on the
nuclear EOS is shown in the left panel of Fig. \ref{mr}. The top and
middle panels give, respectively, the mass of neutron star and its
proton fraction $x_p$, calculated from the MDI interaction with
$x=0, -1$, and $-2$, as functions of its central density. The
symmetry energies obtained from these interactions are shown in the
lower panel together with that from the AV18+$\delta v$+UIX$^{*}$
interaction of Akmal {\it et al.} (APR)~\cite{apr}. It is
interesting to see that up to about $5\rho_0$ the symmetry energy
predicted by APR agrees very well with that from the MDI interaction
with $x=0$. For $x_p$ below 0.14~\cite{lat01}, the direct URCA
process for fast cooling of proto-neutron stars does not proceed
because energy and momentum conservation cannot be simultaneously
satisfied. For EOSs from the MDI interaction with $x=-1$ and $x=-2$,
the condition for direct URCA process is fulfilled for nearly all
neutron stars above 1 \Msun. For the EOS from $x=0$, the minimum
density for the direct URCA process is indicated by the vertical
dotted line, and the corresponding minimum neutron star mass is
indicated by the horizontal dotted line. While this indicates that
neutron stars with masses above 1.39 $\mathrm{M}_{\odot}$ have a
central density above the threshold for the direct Urca process,
this conclusion may be modified by the presence of terms in the
symmetry energy which are quartic in the isospin asymmetry at high
density\cite{andrew06}.

The relations between neutron star masses and radii for above EOSs
are given in the right panel of Fig. \ref{mr}.  Also given are the
constraints due to causality as well as the mass-radius relations
from estimates of the crustal fraction of the moment of inertia
($\Delta I/I=0.014$) in the Vela pulsar~\cite{link} and from the
redshift measurement from Ref.~\cite{cottam}. Allowed equations of
state should lie to the right of the causality line and also cross
the other two lines. The hatched regions are inferred limits on the
radius and the radiation radius (the value of the radius which
observed by an observer at infinity) defined as
$R_{\infty}=R/\sqrt{1-2GM/Rc^2}$ for a 1.4~\Msun~neutron star. It is
seen that the symmetry energy affects strongly the radius of a
neutron star but only slightly its maximum mass~\cite{lat01,pra88}.
These analyses have led to the conclusion that only radii between
11.5 and 13.6 km (or radiation radii between 14.4 and 16.3 km) are
consistent with the EOSs from the MDI interaction with $x=0$ and
$x=-1$ and thus with the terrestrial nuclear laboratory data. The
observational determination of the neutron star radius from the
measured spectral fluxes relies on a numerical model of the neutron
star atmosphere and uses  as inputs the composition of the
atmosphere, a measurement of the distance, the column density of
x-ray absorbing material, and the surface gravitational redshift.
Since many of these quantities are difficult to measure, only a
paucity of radius measurements are available. Nevertheless, it is
interesting to note that the calculations shown in Fig.\ref{mr} are
consistent with recent observations \cite{bob02} including the very
rapidly rotating pulsar discovered recently \cite{rapidns}.
\begin{figure}[th]
\begin{center}
\includegraphics[width=7cm,height=6.cm]{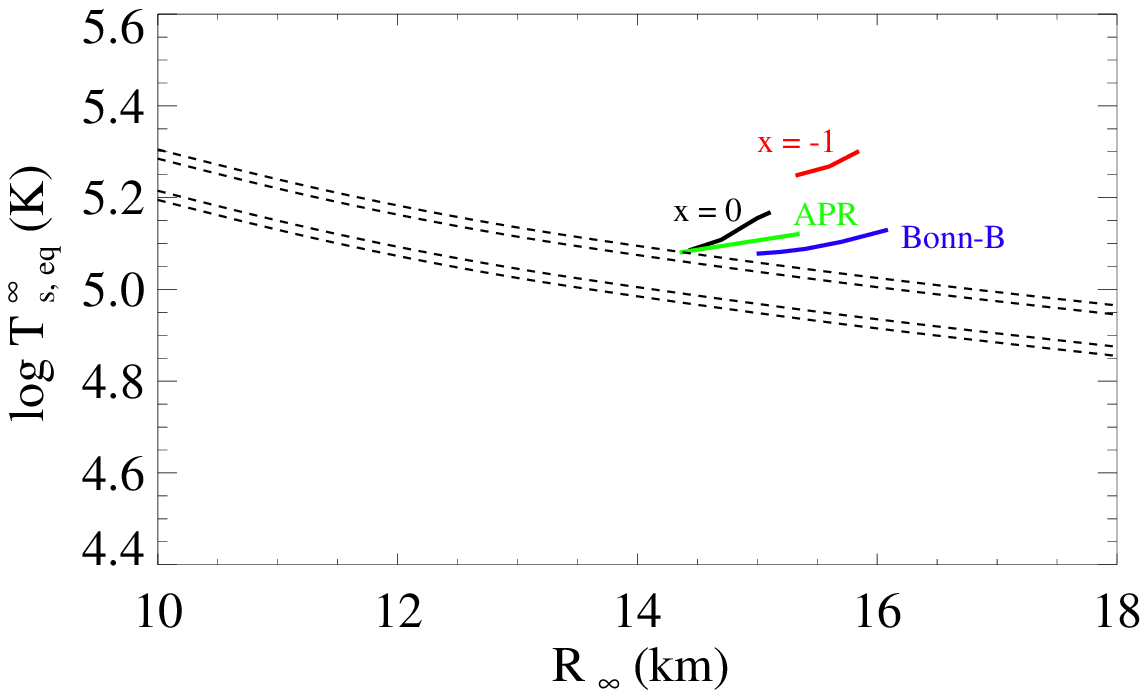}
\includegraphics[width=7cm,height=6.cm]{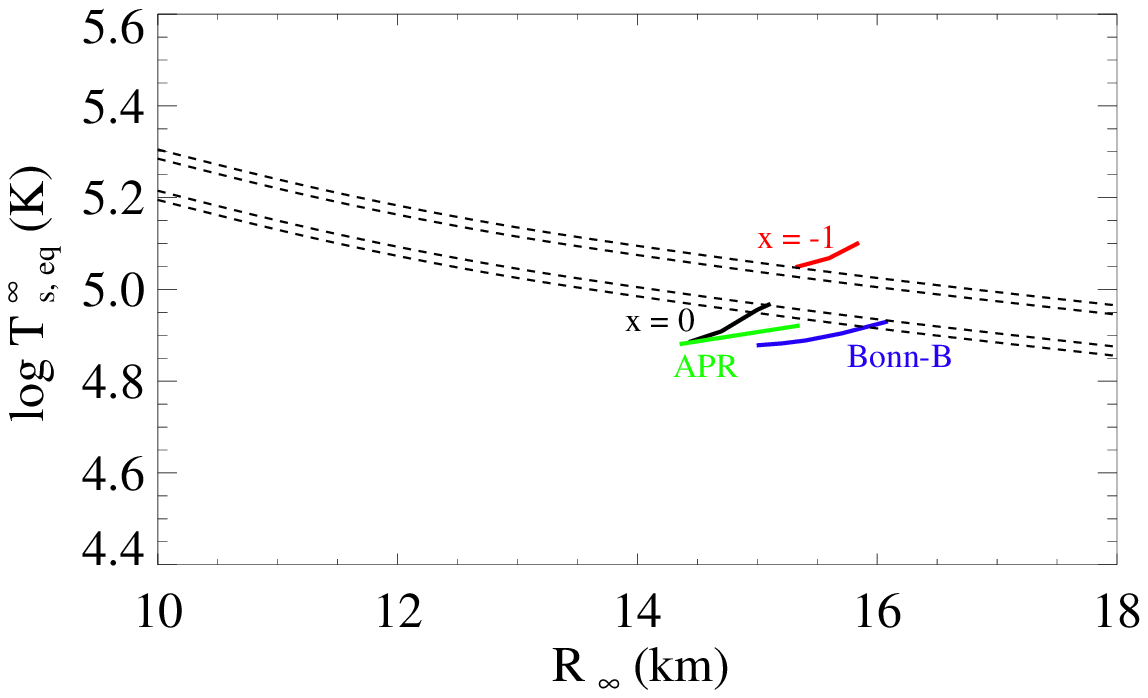}
\caption{{\protect\small Left: Neutron star stationary surface
temperature for stellar models satisfying the mass constraint by van
Straten et al.~\cite{vanStraten:2001zk}. The solid lines are the
predictions versus the stellar radius for the considered neutron
star sequences. Dashed lines correspond to the 68\% and 90\%
confidence contours of the black-body fit of Kargaltsev et
al.~\cite{Kargaltsev:2003eb}. The value of $|\dot{G}/G|=4\times
10^{-12}yr^{-1}$ is chosen so that predictions from the $x=0$ EOS
are just above the observational constraints. Right: Same as left
but assuming $|\dot{G}/G|=8\times 10^{-13}yr^{-1}$.}}\label{g}
\end{center}
\end{figure}
\section{Constraining the changing rate of the gravitational constant G}
Testing the constancy of the gravitational constant G has been a
longstanding fundamental question in natural science. As first
suggested by Jofr\'{e}, Reisenegger and
Fern\'{a}ndez~\cite{Jofre:2006ug}, Dirac's hypothesis \cite{Dirac}
of a decreasing gravitational constant $G$ with time due to the
expansion of the Universe would induce changes in the composition of
neutron stars, causing dissipation and internal heating. Eventually,
neutron stars reach their quasi-stationary states where cooling due
to neutrino and photon emissions balances the internal heating. As
shown in ref.~\cite{Jofre:2006ug} the stationary surface temperature
is directly related to the relative changing rate of G via $
T_s^{\infty}=\tilde{\cal D}\left|\frac{\dot{G}}{G}\right|^{2/7}$,
where the function $\tilde{\cal D}$ is a quantity depending only on
the stellar model and the equation of state. The correlation of
surface temperatures and radii of some old neutron stars may thus
carry useful information about the changing rate of G. Using the
constrained symmetry energy with $x=0$ and $x=-1$ shown in
Fig.\ref{esym}, within the gravitochemical heating formalism, as
shown in Fig.\ref{g} we obtained an upper limit of the relative
changing rate of $|\dot{G}/G|\le4\times 10^{-12}yr^{-1}$. This is
the best available estimate in the literature\cite{kl}. For a
comparison, results with the EOS from the recent
Dirac-Brueckner-Hartree-Fock (DBHF)~\cite{KS1} calculations using
the Bonn B potential are also shown. The Bonn B potential gives
roughly the same result on the stationary surface temperature but
slightly larger radius compared to the $x=0$ case.

\section{Summary}
In summary, the symmetry energy of neutron-rich matter is
fundamentally important for both nuclear physics and astrophysics.
Available data from heavy-ion reactions in terrestrial laboratories
allowed us to constrain the symmetry energy at sub-saturation
densities. This has led to stringent constraints on not only the
nuclear effective interactions but also the cooling mechanisms,
radii of neutron stars and the changing rate of the gravitational
constant G.

\section*{Acknowledgements}
This work was supported in part by the US National Science
Foundation under Grant Nos. PHY-0652548, PHY-0457265, the Research
Corporation, the Welch Foundation under Grant No. A-1358, the Joint
Institute for Nuclear Astrophysics under NSF-PFC grant PHY~02-16783
and the U.S. Department of Energy at Los Alamos National Laboratory
under Contract No. DE-AC52-06NA25396,  the National Natural Science
Foundation of China under Grant Nos. 10575071, and 10675082, MOE of
China under project NCET-05-0392, Shanghai Rising-Star Program under
Grant No. 06QA14024, the SRF for ROCS, SEM of China.

\end{document}